\documentclass[%
 reprint,
 showpacs,preprintnumbers,
 amsmath,amssymb,
 aps,
 pra,
]{revtex4-1}

\usepackage{graphicx} % Include figure files
\usepackage{dcolumn} % Align table columns on decimal point
\usepackage{bm} % bold math
\usepackage{color}
\pdfpageattr {/Group << /S /Transparency /I true /CS /DeviceRGB>>} % fixes adobe reader colorspace problem with transparent pages

\begin{document}

%\preprint{APS/123-QED}

\title{High-resolution spectroscopy of single $\rm Pr^{3+}$ ions on the \mbox{$ \rm ^{3}H_4 -$$\rm^{1}D_2$}
 transition}

\author{Emanuel Eichhammer}
\affiliation{Max Planck Institute for the Science of Light and Friedrich-Alexander-Universit\"at Erlangen-N\"urnberg (FAU),
D-91058 Erlangen, Germany}
\author{Tobias Utikal}
\affiliation{Max Planck Institute for the Science of Light and Friedrich-Alexander-Universit\"at Erlangen-N\"urnberg (FAU),
D-91058 Erlangen, Germany}
\author{Stephan G\"otzinger}
\affiliation{Max Planck Institute for the Science of Light and Friedrich-Alexander-Universit\"at Erlangen-N\"urnberg (FAU),
D-91058 Erlangen, Germany}
\author{Vahid Sandoghdar}
\email{vahid.sandoghdar@mpl.mpg.de}
\affiliation{Max Planck Institute for the Science of Light and Friedrich-Alexander-Universit\"at Erlangen-N\"urnberg (FAU),
D-91058 Erlangen, Germany}

\date{\today}

\begin{abstract}

Rare earth ions in crystals exhibit narrow spectral features and hyperfine-split ground states with exceptionally long coherence times. These features make them ideal platforms for quantum information processing in the solid state. Recently, we reported on the first high-resolution spectroscopy of single $\rm Pr^{3+}$ ions in yttrium orthosilicate (YSO) nanocrystals. While in that work we examined the less explored \mbox{$\rm ^{3}H_4-$$\rm^{3}P_0$} transition at a wavelength of $488\,$nm, here we extend our investigations to the \mbox{$ \rm ^{3}H_4 -$$\rm^{1}D_2$} transition at $606\,$nm. In addition, we present measurements of the second-order autocorrelation function, fluorescence lifetime, and emission spectra of single ions as well as their polarization dependencies on both transitions; these data were not within the reach of the first experiments reported earlier. Furthermore, we show that by a proper choice of the crystallite, one can obtain narrower spectral lines and, thus, resolve the hyperfine levels of the excited state. We expect our results to make single-ion spectroscopy accessible to a larger scientific community.

\end{abstract}


% Quantum Information 
% Optical implementations of quantum information processing and transfer  	
% Quantum optics
% Defects and impurities in crystals; microstructure

%\keywords{Suggested keywords}%Use showkeys class option if keyword display desired
\maketitle

\section{Introduction}

Rare earth ions are ubiquitous in many technologies such as solid-state lasers, amplifiers for optical telecommunication, and magnetic materials. They have also played a central role in the development of high-resolution laser spectroscopy methods such as hole burning and photon echo~\cite{Liu:05}. Transitions to the lowest lying excited states of an isolated rare earth ion take place within its 4f-shell and are dipole-forbidden. In a solid matrix, however, these transitions become weakly probable with long excited-state lifetimes up to milliseconds. Furthermore, because the 4f-intrashell electrons are well shielded by the outer electrons, they are less sensitive against the environment resulting in long spin coherence times of the order of hours~\cite{Zhong:15}. In addition, they possess transitions at various wavelengths in the visible and near infrared, where optical detectors are very sensitive. Furthermore, they are optimally photostable even at room temperature and high excitation powers.

The combination of the above-mentioned features makes rare earth ions also very appealing for emerging applications in quantum information processing. Here, one would often like to access quantum states of well-defined spin via optical interactions and have the possibility of storing qubits for long times. In fact, a number of interesting effects have recently been demonstrated in rare earth-doped bulk crystals such as the robust adiabatic population transfer~\cite{Klein:07}, storage of light in a spin-wave using the atomic frequency comb technique~\cite{Afzelius:10} and storing images in an ion ensemble by electromagnetically induced transparency for one minute~\cite{Heinze:13}. 

For the ultimate control of quantum states at the levels of single material and light particles, however, it is desirable to detect and manipulate single rare earth ions. However, although rare earth ions were on the wish list of single particle microscopy and spectroscopy since the mid 1980s, this goal was only very recently achieved~\cite{Kolesov:12, Yin:13, Kolesov:13, Utikal:14, Siyushev:14, Nakamura:14}.

\section{Brief review of our previous results and new narrower spectra in microcrystals}

\begin{figure}
\includegraphics{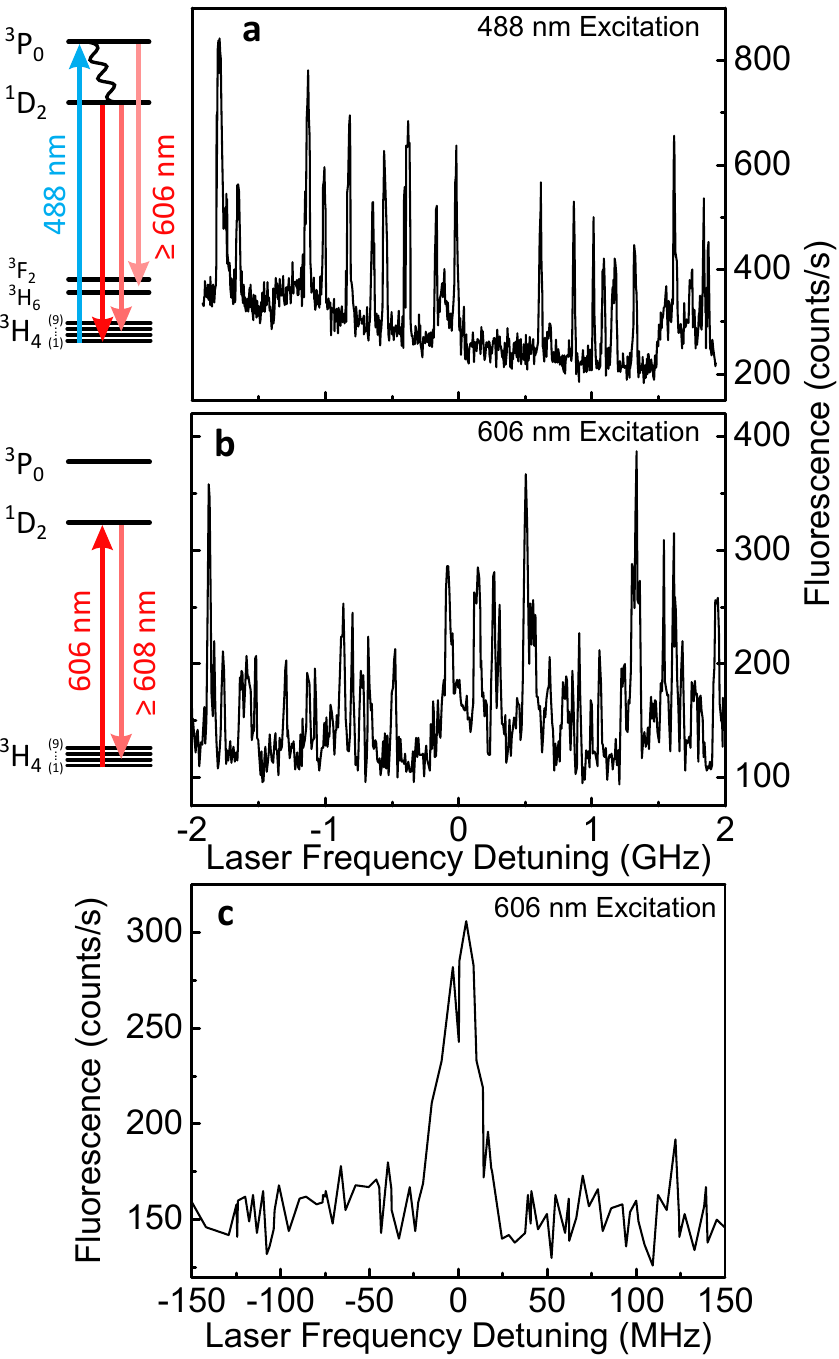}
\caption{\label{fig:scans}(Color online) (a) Fluorescence excitation spectra of ions addressed via the \mbox{$\rm ^{3}H_4-$$\rm^{3}P_0$} (blue) transition at $488\,$nm and the energy level scheme. The scan speed was $4\,$MHz/s. Within our detection spectral range, the $^3$P$_0$ state can directly decay into $^3$H$_6$ and $^3$F$_2$ levels or relax via the $^1$D$_2$ state, which decays into the crystal field levels of the $^3$H$_4$ ground state. (b) Same as in (a) but with the excitation laser tuned to the  \mbox{$^3$H$_4-^1$D$_2$} transition from the lowest crystal field level at $606\,$nm. The scan speed was $5\,$MHz/s. The measured fluorescence is collected at $> 608\,$nm which corresponds to a decay into higher crystal field levels of $^3$H$_4$. The polarization of the excitation light was rotated by $90^\circ$ with respect to the previous blue excitation. (c) Excitation spectrum of a single ion. The three hyperfine levels of the mbox{$^3$H$_4-^1$D$_2$} state separated by 4.6\,MHz and 4.8\,MHz are not resolved.}
\end{figure}

Room-temperature detection of single emitters in the condensed phase requires extremely low concentrations so that neighboring particles are separated by more than the spatial resolution of the imaging system. Under cryogenic conditions, however, one can address individual emitters by tuning the frequency of a narrow-band laser through the inhomogeneous band of the sample even if the concentration is higher by a few orders of magnitude. In this scheme, every time an emitter becomes resonant with the excitation light, it fluoresces. This signal can be detected on a low background because the other emitters in the excitation volume do not absorb light at that frequency. In our previous work~\cite{Utikal:14}, we employed this method pioneered by W.E. Moerner and M. Orrit for the detection of single molecules~\cite{Moerner:89, Orrit:90} to detect single Pr$^{3+}$ ions in Y$_2$SiO$_5$ (YSO). As illustrated in the energy level scheme of Fig.~\ref{fig:scans}(a), we used the \mbox{$\rm ^{3}H_4-$$\rm^{3}P_0$} transition at a wavelength of 488\,nm (blue). The figure shows an example of the excitation spectra of several individual ions over 4 GHz. 

Because of a limited access to samples with suitable concentration, in that work we used YSO nanocrystals to keep the background fluorescence to a manageable level. The spectral diffusion caused by the cracks and surface defects in these samples resulted in the broadening of the resonances from the lifetime-limited linewidth of 82 kHz to several MHz. As a result, the hyperfine levels of the excited state could not be resolved. Furthermore, the weak signal of only tens of photon counts per second prevented us from performing various studies, which we now present in this article. In our current work we have increased our signal level beyond 500 counts per second (see Fig.~\ref{fig:scans}(a)) by using a single photon counting module (SPCM) with a higher quantum efficiency. This leads to faster measurement times and a $1.8$-fold improvement of the signal-to-noise ratio (SNR).

\begin{figure}
\includegraphics[width=7.5cm]{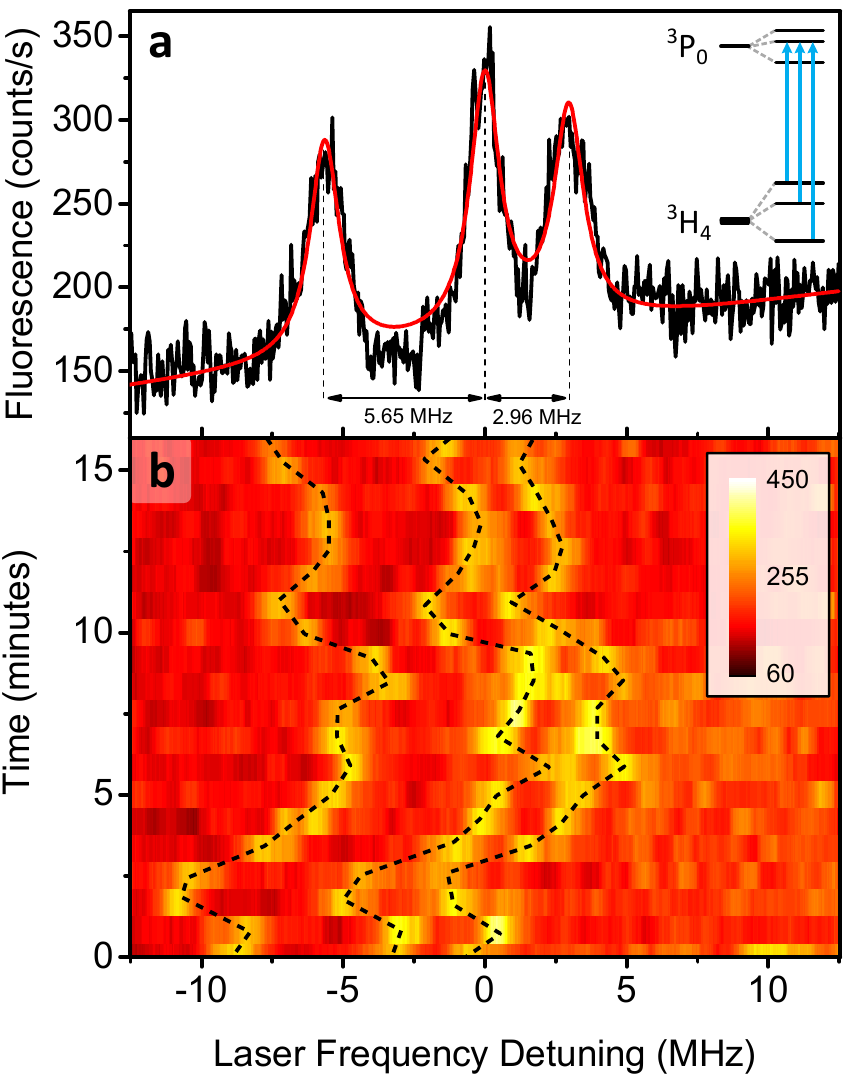}
\caption{\label{fig:narrow-ion}(Color online) (a) An excitation spectrum of a single ion resolving the three hyperfine levels of the $^3$P$_0$ state. This spectrum is the average of the individual spectra shown in (b). The fit (red curve) accounts for a linear background component and yields a linewidth of $1.26\,$MHz. The splitting ordering of $-5.65\,$MHz and $+2.96\,$MHz confirms our previous assignment~\cite{Utikal:14}. (b) Twenty consecutive traces recorded with an integration time of $0.1\,$s per frequency sample. To arrive at the spectrum in (a), the scans were manually aligned in frequency to compensate for spectral diffusion between scans. The color scale shows the recorded counts per second.}
\end{figure}

We now provide new spectra recorded in larger YSO crystallites (diameter 1 to $5\,\mu$m) with a praseodymium doping level of 0.0001\%, which is about 50 times lower than that of the nanocrystals used earlier. As sketched in the inset of Fig.~\ref{fig:narrow-ion}(a), the frequencies of three laser beams detuned by the energy differences of the ground-state hyperfine levels are scanned synchronously. The three peaks in Fig.~\ref{fig:narrow-ion}(a) show that the hyperfine levels of the $^3$P$_0$ state of a single ion are resolved. The measured separations of 5.65 and 2.96\,MHz are in very good agreement with bulk hole-burning spectra presented in our previous publication~\cite{Utikal:14}. A fit to the spectrum composed of three Lorentzian functions yields a transition linewidth of 1.3\,MHz, which is considerably smaller than what we obtained in our former study~\cite{Utikal:14}. 

Interestingly, Fig.~\ref{fig:narrow-ion}(b) reveals that although the fast components of the spectral diffusion have been reduced, a certain level of slow diffusion remains on the time scale of minutes. The fact that the dynamics of transitions to all three hyperfine levels are correlated indicates that the diffusion is not caused by spin instabilities. In future experiments, we expect to eliminate these fully by adjusting the ion doping levels and further control of the crystal size and quality. 

\section{Detection of single ions on the \mbox{$ \rm ^{3}H_4 -$$\rm^{1}D_2$} transition} 

The great majority of bulk studies on Pr:YSO have been reported for the \mbox{$ \rm ^{3}H_4 -$$\rm^{1}D_2$} transition (see the level scheme in Fig.~\ref{fig:scans}(b)) at a wavelength of 606\,nm (red) with a natural linewidth of about 1 kHz~\cite{Equall:95, Nilsson:04, Fraval:05, Klein:07, Afzelius:10, Heinze:13}. Our original motivation for working on the \mbox{$\rm ^{3}H_4-$$\rm^{3}P_0$} transition was the shorter fluorescence lifetime of the $\rm ^{3}P_0$ state and, thus, a broader transition, which makes the requirements on the laser linewidth less stringent. Considering that our current state of the art in linewidth is limited to about 1\,MHz by spectral diffusion, we decided to set up a new dye laser system to see whether it would also be possible to detect single ions via the \mbox{$ \rm ^{3}H_4 -$$\rm^{1}D_2$} transition. 

As in the case of Ref.~\cite{Utikal:14}, sidebands at $-10.19\,$MHz and $+17.3\,$MHz were added to the center excitation laser frequency to prevent shelving of the population in other hyperfine levels of the ground state. Furthermore, we employed a longpass filter designed for $609\,$nm with a $4\,$nm edge to suppress the excitation light. Figure~\ref{fig:scans}(b) shows an example of sharp resonances obtained when sweeping the frequency of the excitation laser over 4 GHz, whereas Fig.~\ref{fig:scans}(c) displays a zoom of the excitation spectrum of a single ion. These measurements were performed in the very same nanocrystal that was used for our previous work~\cite{Utikal:14}. As a result, the hyperfine levels of the \mbox{$ \rm ^{3}H_4 -$$\rm^{1}D_2$} state separated by 4.6\,MHz and 4.8\,MHz are not resolved~\cite{Longdell:02}.

\section{Antibunching and fluorescence lifetime measurements}

A robust proof that the fluorescence signal at each resonance stems from a single emitter can only be provided by the observation of strong antibunching in the second-order autocorrelation function $G^{(2)}(\tau)$ of the emitted photons. Commonly, the short fluorescence lifetimes of molecules, quantum dots or color centers in the nanosecond regime are comparable with the afterpulsing and deadtimes of SPCMs. To get around this problem, one uses a Hanbury Brown-Twiss configuration and searches for a lack of coincidences on two detectors within short time intervals. The long lifetimes of the excited states in rare earth ions permit, however, to use a single SPCM. 

\begin{figure}
\includegraphics[width=8.8cm]{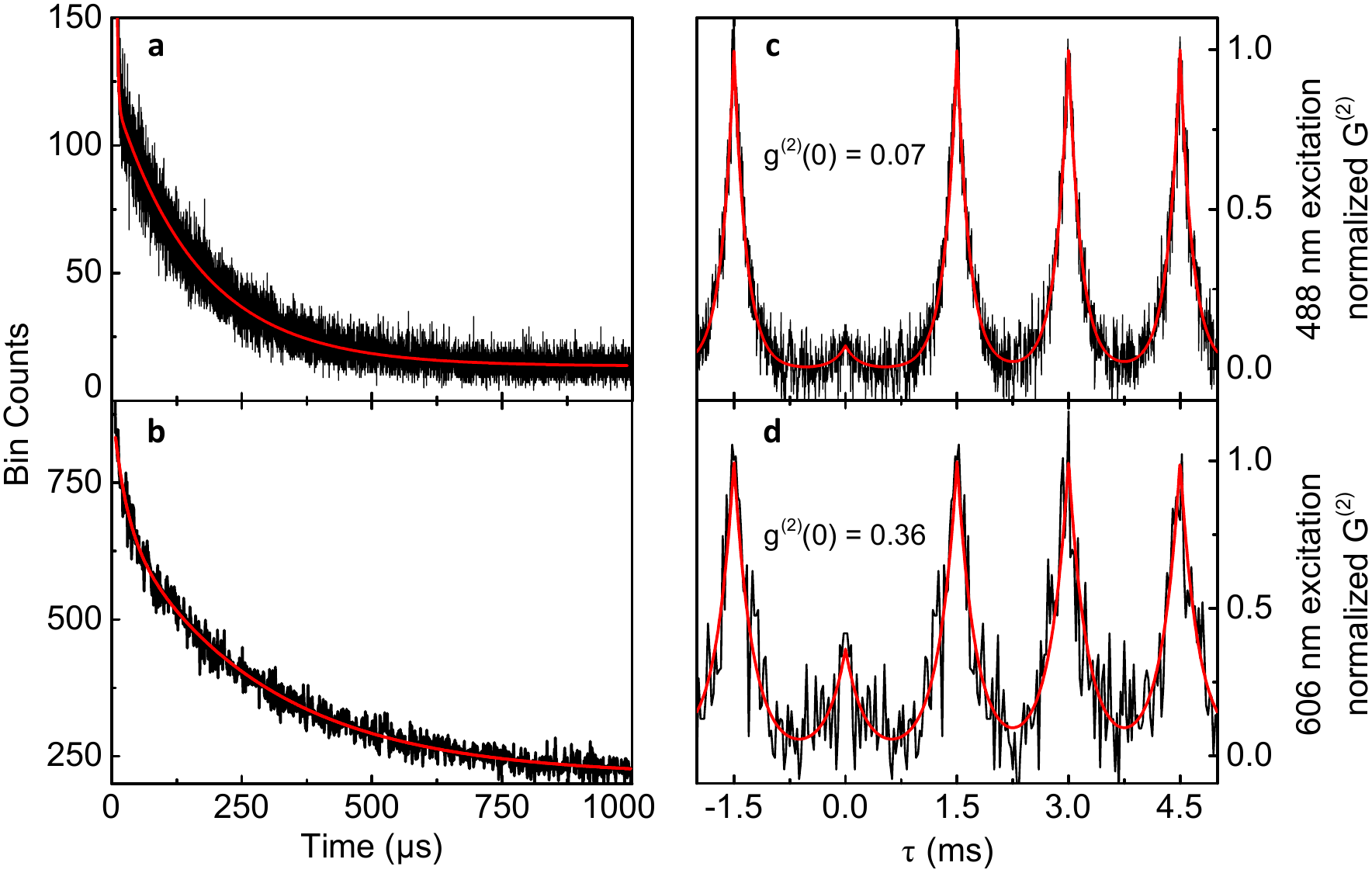}
\caption{\label{fig:photonstatistics}(Color online) (a, b) show the temporal distribution of the photons arrival times from single ions excited at wavelengths of 488\,nm and 606\,nm, respectively. Here the bin sizes were $100\,$ns  and $1\,\mu$s for the blue  and red transition, respectively. (c, d) display the second-order autocorrelation function $G^{(2)}$ normalized by the peak height of single ions studied in (a) and (b), respectively integrated for 2.5 (blue) and 4 (red) hours. Afterpulsing probability was $\approx 0.1\%$. The detector deadtime of $42\,$ns only affects the first bin and was omitted for the analysis. Since all photons in the time tagged photon collection were used to calculate correlation events, the data are symmetric about $\tau = 0\,$ms. The red lines represent fits to the data which are used for normalization and determination of $g^{(2)}(0)$.}
\end{figure}

First, we present measurements of the fluorescence lifetime decay. To obtain these data, we tuned the frequency of the excitation laser beam to a narrow resonance (see Fig.~\ref{fig:scans}) and used an AOM to pulse the excitation light with pulse duration and repetition period of $5\,\mu$s and $1.5\,$ms, respectively. The average laser power was set to achieve 10\% of the continuous-wave fluorescence. The fluorescence was detected with the SPCM and the photons were time-tagged with a time-correlated single photon counting system. Figure~\ref{fig:photonstatistics}(a) displays the fluorescence lifetime curve of the $^3$P$_0$ state from a single ion. The red curve shows a double-exponential fit with $1/e$ times of $2\, \mu$s and $162\, \mu$s, which can be attributed to the decay directly from the $^3$P$_0$ state and indirectly via the $^1$D$_2$ state, respectively. These findings are in good agreement with the results of bulk studies~\cite{Equall:95,Kuleshov:97}. 

Figure~\ref{fig:photonstatistics}(b) shows the same measurement as in (a) but for an ion detected on the red transition. This decay curve was also fitted with two exponentials, yielding $1/e$ times of $25\, \mu$s and $277\, \mu$s although the weight of the contribution with a shorter lifetime only amounted to 4\%. Interestingly, the fluorescence lifetime of $277\, \mu$s for the $^1$D$_2$ state turns out to be 1.7 times longer than the bulk value and our finding in Fig.~\ref{fig:photonstatistics}(a). This difference might be related to the local inhomogeneities of the environment. Access to such variations at the level of individual emitters is one of the fundamental advantages of single-ion spectroscopy. Indeed, we found that the lifetimes vary from ion to ion in the microcrystals. 

To generate clean autocorrelation functions, we removed all photons arriving within $7\,\mu$s of the leading pulse edge and chose time bins of $5\,\mu$s  and $25\,\mu$s for the blue and red transitions, respectively. The center bin at $\tau = 0$ was discarded to remove the majority of the SPCM afterpulses. Figure~\ref{fig:photonstatistics}(c) shows the resulting photon correlation for excitation via the blue transition. The measured histograms were fitted by a piecewise exponential decay and rise function with baseline, overall peak and zero-peak heights as variable parameters. The data were further normalized to set the baseline and the peaks of the fit to 0 and 1, respectively. We note that the data for the blue and the red transitions were not recorded on the same ion (see section~\ref{sec:polarization-dependence}). The measurement on the red channel is somewhat noisier because of a lower fluorescence yield. The resulting values of $g^{(2)}(0) = 0.07$ and 0.36 are well below $0.5$, confirming that in each case the fluorescence stems from a single ion. More importantly, these measurements show that despite the fact that rare earth ions are orders of magnitude weaker than other emitters such as atoms, molecules, quantum dots or color centers, it is possible to study their photon statistics, which is an important quantity for quantum optical investigations.

\section{Polarization dependence} 
\label{sec:polarization-dependence}

Having demonstrated the feasibility of single-ion spectroscopy on the \mbox{$ \rm ^{3}H_4 -$$\rm^{1}D_2$} and \mbox{$\rm ^{3}H_4-$$\rm^{3}P_0$} transitions, we asked whether it was possible to address the same ion through both channels. Here, one might hope to identify the same spectral landscapes (see Fig.~\ref{fig:scans}) within the inhomogeneous band of the two transitions. However, after an extended effort, we found no correlation between them. Furthermore, we found that spectral holes burnt in a bulk sample at 488\,nm could not be addressed by light at 606\,nm and vice versa. We, thus, suspected that each ion experiences different local electromagnetic environments at the two transition wavelengths. Investigations of the polarization behavior of the two excitation channels shed light on this hypothesis. 

\begin{figure}
\includegraphics[width=8cm]{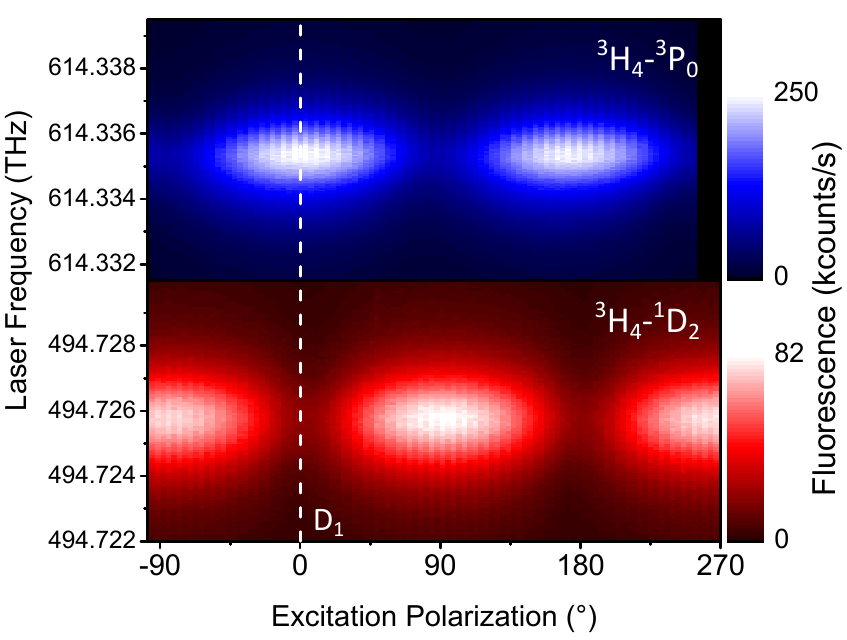}
\caption{\label{fig:polarization}(Color online) Fluorescence from an ensemble of ions as a function of the excitation polarization angle. The top and bottom maps show the signal upon blue and red excitation, respectively. The measurements were performed a few micrometers below the surface of a 0.005\% doped bulk Pr:YSO crystal. In each case, the laser was scanned with $300\,$MHz/s. The zero on the polarization axis was set to align with the D$_1$ crystal axis.}
\end{figure}

The blue (top) and red (bottom) maps in Fig.~\ref{fig:polarization} display the inhomogeneous spectral spread for the transitions \mbox{$\rm ^{3}H_4-$$\rm^{3}P_0$} and \mbox{$ \rm ^{3}H_4 -$$\rm^{1}D_2$} as a function of the excitation polarization. The measurements were performed by focusing the laser beams at a few micrometers below the surface of a bulk crystal without using a solid-immersion lens. We clearly see that the maxima of the two channels are offset by $90^{\circ}$. The orthogonality of the dipole moments associated with the two transitions implies that they are influenced by different local crystal fields and fluctuations. As a result, one cannot expect any correlation between the details of the spectra within the two inhomogeneous distributions. 

\section{Emission spectroscopy}
\label{sec:emission-spectroscopy}

\begin{figure}
\includegraphics[width=8.7cm]{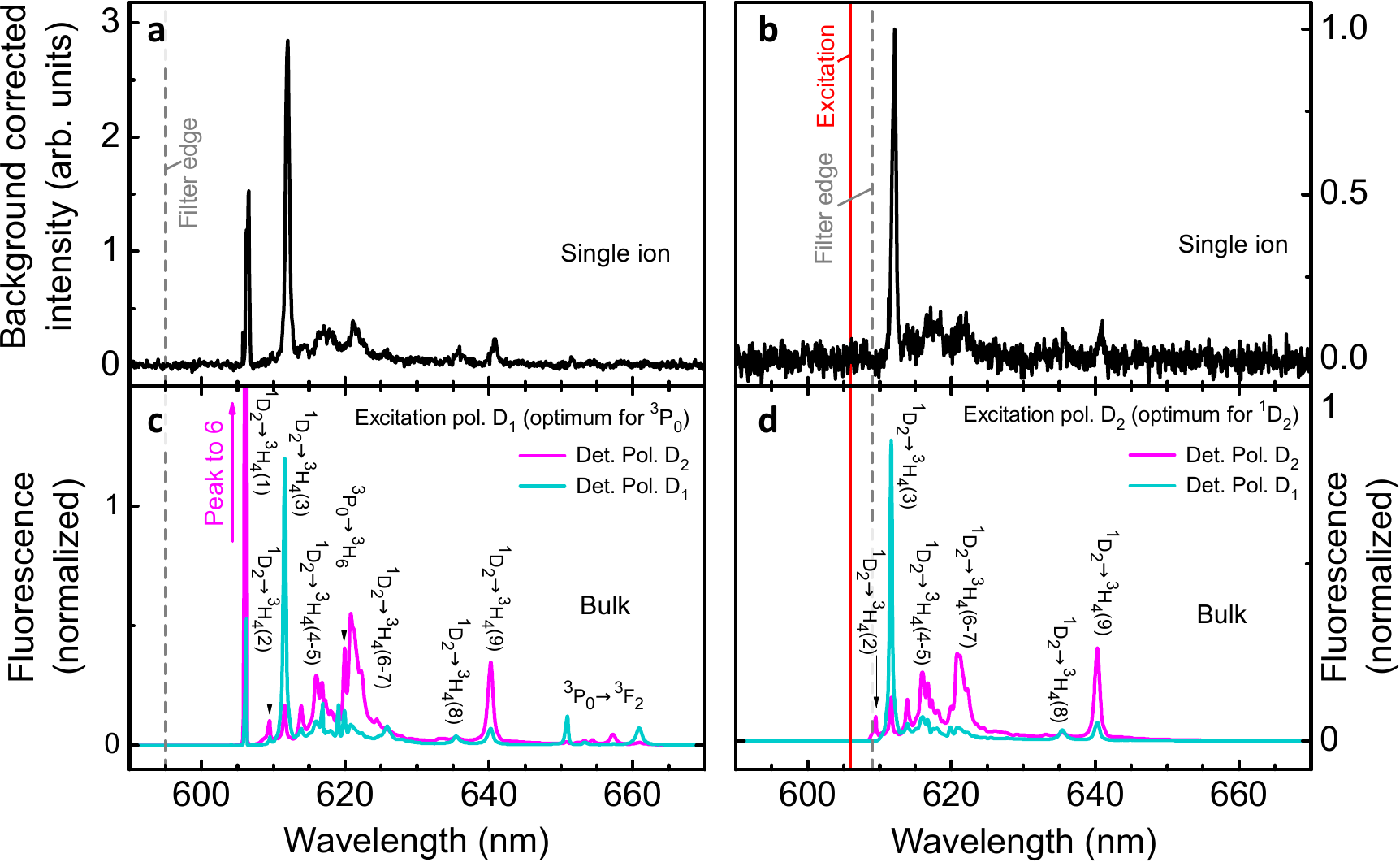}
\caption{\label{fig:spectrum}(Color online) (a, b) The emission spectra of a single ion excited at $488\,$nm and $606\,$nm, respectively. The electron multiplying gain of the EM-CCD camera was set to 200. (b, c) Fluorescence spectra of a bulk Pr:YSO crystal excited at $488\,$nm and $606\,$nm, respectively. Cyan and magenta spectra were recorded with their detection polarizations set along the D$_1$ and D$_2$ crystal axes, respectively.} 
\end{figure}

Next we turn to the emission spectra of single ions. To record such data, we sent the collected fluorescence to a grating spectrometer equipped with a peltier-cooled EM-CCD camera. The spectra were integrated over 2 seconds and accumulated 800 times, whereby we occasionally corrected for drifts between the laser frequency and the ion resonance. The fluorescence background was registered by repeating the measurement with the laser frequency detuned from the ion resonance by 50\,MHz. Figure~\ref{fig:spectrum} shows the single ion spectra for the blue (a) and red (b) transitions. Considering the weak fluorescence rate and the strong dispersion of the emitted photons in a grating spectrometer, such measurements require long integration times and, thus, a high degree of spectral stability. 

To identify the origins of the individual peaks, we have also recorded spectra obtained from a bulk sample (0.005\% Pr-doping). Figure~\ref{fig:spectrum}(c,d) displays these spectra recorded with optimum excitation polarization for the respective transitions (see section~\ref{sec:polarization-dependence}) and detection polarizations along the crystal axes D$_1$ (cyan) and D$_2$ (magenta). Each peak is attributed to a transition in agreement with previously published data \cite{Holliday:93, Equall:95}. We find a clear correspondence between the single-ion and ensemble measurements for each spectral feature although the relative peak heights differ substantially. Here, it is important to note that the emission intensities of different transitions are strongly polarization dependent. In the case of single-ion spectra, the arbitrary orientation of the host crystallite and the strong polarization-dependent coupling of the emission into the solid-immersion lens make it difficult to quantify the relative peak heights.

\section{Conclusion}

In the present work we have extended our first report on the detection and spectroscopy of single praseodymium ions on the \mbox{$\rm ^{3}H_4-$$\rm^{3}P_0$} transition to the \mbox{$ \rm ^{3}H_4 -$$\rm^{1}D_2$} transition, which has been much more widely studied. In addition, we have improved the signal-to-noise ratio of our detection process and demonstrated narrower linewidths than our previous work, allowing us to resolve the hyperfine levels of the $^3$P$_0$ state. Furthermore, we have presented several new measurements that are usually desirable at the single-emitter level. These include  the measurement of the fluorescence lifetime, autocorellation function and strong photon antibunching as well as emission spectra of single ions. Finally, we have examined the polarization dependence of the excitation and emission channels and have provided evidence that a given ion experiences different local fields and frequency shifts on its various transitions. The findings of this work further fuel the recent emergence of activities on the detection and spectroscopy of rare earth systems at the single-ion level~\cite{Kolesov:12, Yin:13, Kolesov:13, Utikal:14, Siyushev:14, Nakamura:14}.

%\bibliography{vahid-bibfile}

\begin{thebibliography}{19}%
\makeatletter
\providecommand \@ifxundefined [1]{%
 \@ifx{#1\undefined}
}%
\providecommand \@ifnum [1]{%
 \ifnum #1\expandafter \@firstoftwo
 \else \expandafter \@secondoftwo
 \fi
}%
\providecommand \@ifx [1]{%
 \ifx #1\expandafter \@firstoftwo
 \else \expandafter \@secondoftwo
 \fi
}%
\providecommand \natexlab [1]{#1}%
\providecommand \enquote  [1]{``#1''}%
\providecommand \bibnamefont  [1]{#1}%
\providecommand \bibfnamefont [1]{#1}%
\providecommand \citenamefont [1]{#1}%
\providecommand \href@noop [0]{\@secondoftwo}%
\providecommand \href [0]{\begingroup \@sanitize@url \@href}%
\providecommand \@href[1]{\@@startlink{#1}\@@href}%
\providecommand \@@href[1]{\endgroup#1\@@endlink}%
\providecommand \@sanitize@url [0]{\catcode `\\12\catcode `\$12\catcode
  `\&12\catcode `\#12\catcode `\^12\catcode `\_12\catcode `\%12\relax}%
\providecommand \@@startlink[1]{}%
\providecommand \@@endlink[0]{}%
\providecommand \url  [0]{\begingroup\@sanitize@url \@url }%
\providecommand \@url [1]{\endgroup\@href {#1}{\urlprefix }}%
\providecommand \urlprefix  [0]{URL }%
\providecommand \Eprint [0]{\href }%
\providecommand \doibase [0]{http://dx.doi.org/}%
\providecommand \selectlanguage [0]{\@gobble}%
\providecommand \bibinfo  [0]{\@secondoftwo}%
\providecommand \bibfield  [0]{\@secondoftwo}%
\providecommand \translation [1]{[#1]}%
\providecommand \BibitemOpen [0]{}%
\providecommand \bibitemStop [0]{}%
\providecommand \bibitemNoStop [0]{.\EOS\space}%
\providecommand \EOS [0]{\spacefactor3000\relax}%
\providecommand \BibitemShut  [1]{\csname bibitem#1\endcsname}%
\let\auto@bib@innerbib\@empty
%</preamble>
\bibitem [{\citenamefont {Liu}\ and\ \citenamefont {Jacquier}(2005)}]{Liu:05}%
  \BibitemOpen
  \bibfield  {author} {\bibinfo {author} {\bibfnamefont {G.}~\bibnamefont
  {Liu}}\ and\ \bibinfo {author} {\bibfnamefont {B.}~\bibnamefont {Jacquier}},\
  }\href@noop {} {\emph {\bibinfo {title} {{Spectroscopic Properties of Rare
  Earths in Optical Materials}}}}\ (\bibinfo  {publisher} {Springer},\ \bibinfo
  {year} {2005})\BibitemShut {NoStop}%
\bibitem [{\citenamefont {Zhong}\ \emph {et~al.}(2015)\citenamefont {Zhong},
  \citenamefont {Hedges}, \citenamefont {Ahlefeldt}, \citenamefont
  {Bartholomew}, \citenamefont {Beavan}, \citenamefont {Wittig}, \citenamefont
  {Longdell},\ and\ \citenamefont {Sellars}}]{Zhong:15}%
  \BibitemOpen
  \bibfield  {author} {\bibinfo {author} {\bibfnamefont {M.}~\bibnamefont
  {Zhong}}, \bibinfo {author} {\bibfnamefont {M.~P.}\ \bibnamefont {Hedges}},
  \bibinfo {author} {\bibfnamefont {R.~L.}\ \bibnamefont {Ahlefeldt}}, \bibinfo
  {author} {\bibfnamefont {J.~G.}\ \bibnamefont {Bartholomew}}, \bibinfo
  {author} {\bibfnamefont {S.~E.}\ \bibnamefont {Beavan}}, \bibinfo {author}
  {\bibfnamefont {S.~M.}\ \bibnamefont {Wittig}}, \bibinfo {author}
  {\bibfnamefont {J.~J.}\ \bibnamefont {Longdell}}, \ and\ \bibinfo {author}
  {\bibfnamefont {M.~J.}\ \bibnamefont {Sellars}},\ }\href@noop {} {\bibfield
  {journal} {\bibinfo  {journal} {Nature}\ }\textbf {\bibinfo {volume} {517}},\
  \bibinfo {pages} {177} (\bibinfo {year} {2015})}\BibitemShut {NoStop}%
\bibitem [{\citenamefont {Klein}\ \emph {et~al.}(2007)\citenamefont {Klein},
  \citenamefont {Beil},\ and\ \citenamefont {Halfmann}}]{Klein:07}%
  \BibitemOpen
  \bibfield  {author} {\bibinfo {author} {\bibfnamefont {J.}~\bibnamefont
  {Klein}}, \bibinfo {author} {\bibfnamefont {F.}~\bibnamefont {Beil}}, \ and\
  \bibinfo {author} {\bibfnamefont {T.}~\bibnamefont {Halfmann}},\ }\href@noop
  {} {\bibfield  {journal} {\bibinfo  {journal} {Phys. Rev. Lett.}\ }\textbf
  {\bibinfo {volume} {99}},\ \bibinfo {pages} {113003} (\bibinfo {year}
  {2007})}\BibitemShut {NoStop}%
\bibitem [{\citenamefont {Afzelius}\ \emph {et~al.}(2010)\citenamefont
  {Afzelius}, \citenamefont {Usmani}, \citenamefont {Amari}, \citenamefont
  {Lauritzen}, \citenamefont {Walther}, \citenamefont {Simon}, \citenamefont
  {Sangouard}, \citenamefont {Min\'a\ifmmode~\check{r}\else \v{r}\fi{}},
  \citenamefont {de~Riedmatten}, \citenamefont {Gisin},\ and\ \citenamefont
  {Kr\"oll}}]{Afzelius:10}%
  \BibitemOpen
  \bibfield  {author} {\bibinfo {author} {\bibfnamefont {M.}~\bibnamefont
  {Afzelius}}, \bibinfo {author} {\bibfnamefont {I.}~\bibnamefont {Usmani}},
  \bibinfo {author} {\bibfnamefont {A.}~\bibnamefont {Amari}}, \bibinfo
  {author} {\bibfnamefont {B.}~\bibnamefont {Lauritzen}}, \bibinfo {author}
  {\bibfnamefont {A.}~\bibnamefont {Walther}}, \bibinfo {author} {\bibfnamefont
  {C.}~\bibnamefont {Simon}}, \bibinfo {author} {\bibfnamefont
  {N.}~\bibnamefont {Sangouard}}, \bibinfo {author} {\bibfnamefont {J.~c.~v.}\
  \bibnamefont {Min\'a\ifmmode~\check{r}\else \v{r}\fi{}}}, \bibinfo {author}
  {\bibfnamefont {H.}~\bibnamefont {de~Riedmatten}}, \bibinfo {author}
  {\bibfnamefont {N.}~\bibnamefont {Gisin}}, \ and\ \bibinfo {author}
  {\bibfnamefont {S.}~\bibnamefont {Kr\"oll}},\ }\href@noop {} {\bibfield
  {journal} {\bibinfo  {journal} {Phys. Rev. Lett.}\ }\textbf {\bibinfo
  {volume} {104}},\ \bibinfo {pages} {040503} (\bibinfo {year}
  {2010})}\BibitemShut {NoStop}%
\bibitem [{\citenamefont {Heinze}\ \emph {et~al.}(2013)\citenamefont {Heinze},
  \citenamefont {Hubrich},\ and\ \citenamefont {Halfmann}}]{Heinze:13}%
  \BibitemOpen
  \bibfield  {author} {\bibinfo {author} {\bibfnamefont {G.}~\bibnamefont
  {Heinze}}, \bibinfo {author} {\bibfnamefont {C.}~\bibnamefont {Hubrich}}, \
  and\ \bibinfo {author} {\bibfnamefont {T.}~\bibnamefont {Halfmann}},\
  }\href@noop {} {\bibfield  {journal} {\bibinfo  {journal} {Phys. Rev. Lett.}\
  }\textbf {\bibinfo {volume} {111}},\ \bibinfo {pages} {033601} (\bibinfo
  {year} {2013})}\BibitemShut {NoStop}%
\bibitem [{\citenamefont {Kolesov}\ \emph {et~al.}(2012)\citenamefont
  {Kolesov}, \citenamefont {Xia}, \citenamefont {Reuter}, \citenamefont
  {St{\"o}hr}, \citenamefont {Zappe}, \citenamefont {Meijer}, \citenamefont
  {Hemmer},\ and\ \citenamefont {Wrachtrup}}]{Kolesov:12}%
  \BibitemOpen
  \bibfield  {author} {\bibinfo {author} {\bibfnamefont {R.}~\bibnamefont
  {Kolesov}}, \bibinfo {author} {\bibfnamefont {K.}~\bibnamefont {Xia}},
  \bibinfo {author} {\bibfnamefont {R.}~\bibnamefont {Reuter}}, \bibinfo
  {author} {\bibfnamefont {R.}~\bibnamefont {St{\"o}hr}}, \bibinfo {author}
  {\bibfnamefont {A.}~\bibnamefont {Zappe}}, \bibinfo {author} {\bibfnamefont
  {J.}~\bibnamefont {Meijer}}, \bibinfo {author} {\bibfnamefont {P.~R.}\
  \bibnamefont {Hemmer}}, \ and\ \bibinfo {author} {\bibfnamefont
  {J.}~\bibnamefont {Wrachtrup}},\ }\href@noop {} {\bibfield  {journal}
  {\bibinfo  {journal} {Nat. Comm.}\ }\textbf {\bibinfo {volume} {3}},\
  \bibinfo {pages} {1029} (\bibinfo {year} {2012})}\BibitemShut {NoStop}%
\bibitem [{\citenamefont {Yin}\ \emph {et~al.}(2013)\citenamefont {Yin},
  \citenamefont {Rancic}, \citenamefont {de~Boo}, \citenamefont {Stavrias},
  \citenamefont {{McCallum}}, \citenamefont {Sellars},\ and\ \citenamefont
  {Rogge}}]{Yin:13}%
  \BibitemOpen
  \bibfield  {author} {\bibinfo {author} {\bibfnamefont {C.}~\bibnamefont
  {Yin}}, \bibinfo {author} {\bibfnamefont {M.}~\bibnamefont {Rancic}},
  \bibinfo {author} {\bibfnamefont {G.~G.}\ \bibnamefont {de~Boo}}, \bibinfo
  {author} {\bibfnamefont {N.}~\bibnamefont {Stavrias}}, \bibinfo {author}
  {\bibfnamefont {J.~C.}\ \bibnamefont {{McCallum}}}, \bibinfo {author}
  {\bibfnamefont {M.~J.}\ \bibnamefont {Sellars}}, \ and\ \bibinfo {author}
  {\bibfnamefont {S.}~\bibnamefont {Rogge}},\ }\href@noop {} {\bibfield
  {journal} {\bibinfo  {journal} {Nature}\ }\textbf {\bibinfo {volume} {497}},\
  \bibinfo {pages} {91} (\bibinfo {year} {2013})}\BibitemShut {NoStop}%
\bibitem [{\citenamefont {Kolesov}\ \emph {et~al.}(2013)\citenamefont
  {Kolesov}, \citenamefont {Xia}, \citenamefont {Reuter}, \citenamefont
  {Jamali}, \citenamefont {St{\"o}hr}, \citenamefont {Inal}, \citenamefont
  {Siyushev},\ and\ \citenamefont {Wrachtrup}}]{Kolesov:13}%
  \BibitemOpen
  \bibfield  {author} {\bibinfo {author} {\bibfnamefont {R.}~\bibnamefont
  {Kolesov}}, \bibinfo {author} {\bibfnamefont {K.}~\bibnamefont {Xia}},
  \bibinfo {author} {\bibfnamefont {R.}~\bibnamefont {Reuter}}, \bibinfo
  {author} {\bibfnamefont {M.}~\bibnamefont {Jamali}}, \bibinfo {author}
  {\bibfnamefont {R.}~\bibnamefont {St{\"o}hr}}, \bibinfo {author}
  {\bibfnamefont {T.}~\bibnamefont {Inal}}, \bibinfo {author} {\bibfnamefont
  {P.}~\bibnamefont {Siyushev}}, \ and\ \bibinfo {author} {\bibfnamefont
  {J.}~\bibnamefont {Wrachtrup}},\ }\href@noop {} {\bibfield  {journal}
  {\bibinfo  {journal} {Phys. Rev. Lett.}\ }\textbf {\bibinfo {volume} {111}},\
  \bibinfo {pages} {120502} (\bibinfo {year} {2013})}\BibitemShut {NoStop}%
\bibitem [{\citenamefont {Utikal}\ \emph {et~al.}(2014)\citenamefont {Utikal},
  \citenamefont {Eichhammer}, \citenamefont {Petersen}, \citenamefont {Renn},
  \citenamefont {G\"otzinger},\ and\ \citenamefont {Sandoghdar}}]{Utikal:14}%
  \BibitemOpen
  \bibfield  {author} {\bibinfo {author} {\bibfnamefont {T.}~\bibnamefont
  {Utikal}}, \bibinfo {author} {\bibfnamefont {E.}~\bibnamefont {Eichhammer}},
  \bibinfo {author} {\bibfnamefont {L.}~\bibnamefont {Petersen}}, \bibinfo
  {author} {\bibfnamefont {A.}~\bibnamefont {Renn}}, \bibinfo {author}
  {\bibfnamefont {S.}~\bibnamefont {G\"otzinger}}, \ and\ \bibinfo {author}
  {\bibfnamefont {V.}~\bibnamefont {Sandoghdar}},\ }\href@noop {} {\bibfield
  {journal} {\bibinfo  {journal} {Nat. Comm.}\ }\textbf {\bibinfo {volume}
  {5}},\ \bibinfo {pages} {3627} (\bibinfo {year} {2014})}\BibitemShut
  {NoStop}%
\bibitem [{\citenamefont {Siyushev}\ \emph {et~al.}(2014)\citenamefont
  {Siyushev}, \citenamefont {Xia}, \citenamefont {Reuter}, \citenamefont
  {Jamali}, \citenamefont {Zhao}, \citenamefont {Yang}, \citenamefont {Duan},
  \citenamefont {Kukharchyk}, \citenamefont {Wieck}, \citenamefont {Kolesov},\
  and\ \citenamefont {Wrachtrup}}]{Siyushev:14}%
  \BibitemOpen
  \bibfield  {author} {\bibinfo {author} {\bibfnamefont {P.}~\bibnamefont
  {Siyushev}}, \bibinfo {author} {\bibfnamefont {K.}~\bibnamefont {Xia}},
  \bibinfo {author} {\bibfnamefont {R.}~\bibnamefont {Reuter}}, \bibinfo
  {author} {\bibfnamefont {M.}~\bibnamefont {Jamali}}, \bibinfo {author}
  {\bibfnamefont {N.}~\bibnamefont {Zhao}}, \bibinfo {author} {\bibfnamefont
  {N.}~\bibnamefont {Yang}}, \bibinfo {author} {\bibfnamefont {C.}~\bibnamefont
  {Duan}}, \bibinfo {author} {\bibfnamefont {N.}~\bibnamefont {Kukharchyk}},
  \bibinfo {author} {\bibfnamefont {A.~D.}\ \bibnamefont {Wieck}}, \bibinfo
  {author} {\bibfnamefont {R.}~\bibnamefont {Kolesov}}, \ and\ \bibinfo
  {author} {\bibfnamefont {J.}~\bibnamefont {Wrachtrup}},\ }\href@noop {}
  {\bibfield  {journal} {\bibinfo  {journal} {Nat. Comm.}\ }\textbf {\bibinfo
  {volume} {5}} (\bibinfo {year} {2014})}\BibitemShut {NoStop}%
\bibitem [{\citenamefont {Nakamura}\ \emph {et~al.}(2014)\citenamefont
  {Nakamura}, \citenamefont {Yoshihiro}, \citenamefont {Inagawa}, \citenamefont
  {Fujiyoshi},\ and\ \citenamefont {Matsushita}}]{Nakamura:14}%
  \BibitemOpen
  \bibfield  {author} {\bibinfo {author} {\bibfnamefont {I.}~\bibnamefont
  {Nakamura}}, \bibinfo {author} {\bibfnamefont {T.}~\bibnamefont {Yoshihiro}},
  \bibinfo {author} {\bibfnamefont {H.}~\bibnamefont {Inagawa}}, \bibinfo
  {author} {\bibfnamefont {S.}~\bibnamefont {Fujiyoshi}}, \ and\ \bibinfo
  {author} {\bibfnamefont {M.}~\bibnamefont {Matsushita}},\ }\href@noop {}
  {\bibfield  {journal} {\bibinfo  {journal} {Scientific Rep.}\ }\textbf
  {\bibinfo {volume} {4}},\ \bibinfo {pages} {7364} (\bibinfo {year}
  {2014})}\BibitemShut {NoStop}%
\bibitem [{\citenamefont {Moerner}\ and\ \citenamefont
  {Kador}(1989)}]{Moerner:89}%
  \BibitemOpen
  \bibfield  {author} {\bibinfo {author} {\bibfnamefont {W.~E.}\ \bibnamefont
  {Moerner}}\ and\ \bibinfo {author} {\bibfnamefont {L.}~\bibnamefont
  {Kador}},\ }\href@noop {} {\bibfield  {journal} {\bibinfo  {journal} {Phys.
  Rev. Lett.}\ }\textbf {\bibinfo {volume} {62}},\ \bibinfo {pages} {2535}
  (\bibinfo {year} {1989})}\BibitemShut {NoStop}%
\bibitem [{\citenamefont {Orrit}\ and\ \citenamefont
  {Bernard}(1990)}]{Orrit:90}%
  \BibitemOpen
  \bibfield  {author} {\bibinfo {author} {\bibfnamefont {M.}~\bibnamefont
  {Orrit}}\ and\ \bibinfo {author} {\bibfnamefont {J.}~\bibnamefont
  {Bernard}},\ }\href@noop {} {\bibfield  {journal} {\bibinfo  {journal} {Phys.
  Rev. Lett.}\ }\textbf {\bibinfo {volume} {65}},\ \bibinfo {pages} {2716}
  (\bibinfo {year} {1990})}\BibitemShut {NoStop}%
\bibitem [{\citenamefont {Equall}\ \emph {et~al.}(1995)\citenamefont {Equall},
  \citenamefont {Cone},\ and\ \citenamefont {Macfarlane}}]{Equall:95}%
  \BibitemOpen
  \bibfield  {author} {\bibinfo {author} {\bibfnamefont {R.~W.}\ \bibnamefont
  {Equall}}, \bibinfo {author} {\bibfnamefont {R.~L.}\ \bibnamefont {Cone}}, \
  and\ \bibinfo {author} {\bibfnamefont {R.~M.}\ \bibnamefont {Macfarlane}},\
  }\href@noop {} {\bibfield  {journal} {\bibinfo  {journal} {Phys. Rev. B}\
  }\textbf {\bibinfo {volume} {52}},\ \bibinfo {pages} {3963} (\bibinfo {year}
  {1995})}\BibitemShut {NoStop}%
\bibitem [{\citenamefont {Nilsson}\ \emph {et~al.}(2004)\citenamefont
  {Nilsson}, \citenamefont {Rippe}, \citenamefont {Kr{\"o}ll}, \citenamefont
  {Klieber},\ and\ \citenamefont {Suter}}]{Nilsson:04}%
  \BibitemOpen
  \bibfield  {author} {\bibinfo {author} {\bibfnamefont {M.}~\bibnamefont
  {Nilsson}}, \bibinfo {author} {\bibfnamefont {L.}~\bibnamefont {Rippe}},
  \bibinfo {author} {\bibfnamefont {S.}~\bibnamefont {Kr{\"o}ll}}, \bibinfo
  {author} {\bibfnamefont {R.}~\bibnamefont {Klieber}}, \ and\ \bibinfo
  {author} {\bibfnamefont {D.}~\bibnamefont {Suter}},\ }\href@noop {}
  {\bibfield  {journal} {\bibinfo  {journal} {Phys. Rev. B}\ }\textbf {\bibinfo
  {volume} {70}},\ \bibinfo {pages} {214116} (\bibinfo {year}
  {2004})}\BibitemShut {NoStop}%
\bibitem [{\citenamefont {Fraval}\ \emph {et~al.}(2005)\citenamefont {Fraval},
  \citenamefont {Sellars},\ and\ \citenamefont {Longdell}}]{Fraval:05}%
  \BibitemOpen
  \bibfield  {author} {\bibinfo {author} {\bibfnamefont {E.}~\bibnamefont
  {Fraval}}, \bibinfo {author} {\bibfnamefont {M.~J.}\ \bibnamefont {Sellars}},
  \ and\ \bibinfo {author} {\bibfnamefont {J.~J.}\ \bibnamefont {Longdell}},\
  }\href@noop {} {\bibfield  {journal} {\bibinfo  {journal} {Phys. Rev. Lett.}\
  }\textbf {\bibinfo {volume} {95}},\ \bibinfo {pages} {030506} (\bibinfo
  {year} {2005})}\BibitemShut {NoStop}%
\bibitem [{\citenamefont {Longdell}\ \emph {et~al.}(2002)\citenamefont
  {Longdell}, \citenamefont {Sellars},\ and\ \citenamefont
  {Manson}}]{Longdell:02}%
  \BibitemOpen
  \bibfield  {author} {\bibinfo {author} {\bibfnamefont {J.~J.}\ \bibnamefont
  {Longdell}}, \bibinfo {author} {\bibfnamefont {M.~J.}\ \bibnamefont
  {Sellars}}, \ and\ \bibinfo {author} {\bibfnamefont {N.~B.}\ \bibnamefont
  {Manson}},\ }\href@noop {} {\bibfield  {journal} {\bibinfo  {journal} {Phys.
  Rev. B.}\ }\textbf {\bibinfo {volume} {66}},\ \bibinfo {pages} {035101}
  (\bibinfo {year} {2002})}\BibitemShut {NoStop}%
\bibitem [{\citenamefont {Kuleshov}\ \emph {et~al.}(1997)\citenamefont
  {Kuleshov}, \citenamefont {Shcherbitsky}, \citenamefont {Lagatsky},
  \citenamefont {Mikhailov}, \citenamefont {Minkov}, \citenamefont {Danger},
  \citenamefont {Sandrock},\ and\ \citenamefont {Huber}}]{Kuleshov:97}%
  \BibitemOpen
  \bibfield  {author} {\bibinfo {author} {\bibfnamefont {N.~V.}\ \bibnamefont
  {Kuleshov}}, \bibinfo {author} {\bibfnamefont {V.~G.}\ \bibnamefont
  {Shcherbitsky}}, \bibinfo {author} {\bibfnamefont {A.~A.}\ \bibnamefont
  {Lagatsky}}, \bibinfo {author} {\bibfnamefont {V.~P.}\ \bibnamefont
  {Mikhailov}}, \bibinfo {author} {\bibfnamefont {B.~I.}\ \bibnamefont
  {Minkov}}, \bibinfo {author} {\bibfnamefont {T.}~\bibnamefont {Danger}},
  \bibinfo {author} {\bibfnamefont {T.}~\bibnamefont {Sandrock}}, \ and\
  \bibinfo {author} {\bibfnamefont {G.}~\bibnamefont {Huber}},\ }\href@noop {}
  {\bibfield  {journal} {\bibinfo  {journal} {J. Lumin.}\ }\textbf {\bibinfo
  {volume} {71}},\ \bibinfo {pages} {27} (\bibinfo {year} {1997})}\BibitemShut
  {NoStop}%
\bibitem [{\citenamefont {Holliday}\ \emph {et~al.}(1993)\citenamefont
  {Holliday}, \citenamefont {Croci}, \citenamefont {Vauthey},\ and\
  \citenamefont {Wild}}]{Holliday:93}%
  \BibitemOpen
  \bibfield  {author} {\bibinfo {author} {\bibfnamefont {K.}~\bibnamefont
  {Holliday}}, \bibinfo {author} {\bibfnamefont {M.}~\bibnamefont {Croci}},
  \bibinfo {author} {\bibfnamefont {E.}~\bibnamefont {Vauthey}}, \ and\
  \bibinfo {author} {\bibfnamefont {U.}~\bibnamefont {Wild}},\ }\href@noop {}
  {\bibfield  {journal} {\bibinfo  {journal} {Phys. Rev. B.}\ }\textbf
  {\bibinfo {volume} {47}},\ \bibinfo {pages} {14741} (\bibinfo {year}
  {1993})}\BibitemShut {NoStop}%
\end{thebibliography}

\end{document}